\documentstyle[11pt,psfig,twoside]{article}

\begin{document}

\newcommand{\dd}{\,{\rm d}}
\newcommand{\ie}{{\it i.e.},\,}
\newcommand{\etal}{{\it et al.\ }}
\newcommand{\eg}{{\it e.g.},\,}
\newcommand{\cf}{{\it cf.\ }}
\newcommand{\vs}{{\it vs.\ }}
\newcommand{\zdot}{\makebox[0pt][l]{.}}
\newcommand{\up}[1]{\ifmmode^{\rm #1}\else$^{\rm #1}$\fi}
\newcommand{\dn}[1]{\ifmmode_{\rm #1}\else$_{\rm #1}$\fi}
\newcommand{\upd}{\up{d}}
\newcommand{\uph}{\up{h}}
\newcommand{\upm}{\up{m}}
\newcommand{\ups}{\up{s}}
\newcommand{\arcd}{\ifmmode^{\circ}\else$^{\circ}$\fi}
\newcommand{\arcm}{\ifmmode{'}\else$'$\fi}
\newcommand{\arcs}{\ifmmode{''}\else$''$\fi}
\newcommand{\MS}{{\rm M}\ifmmode_{\odot}\else$_{\odot}$\fi}
\newcommand{\RS}{{\rm R}\ifmmode_{\odot}\else$_{\odot}$\fi}
\newcommand{\LS}{{\rm L}\ifmmode_{\odot}\else$_{\odot}$\fi}

\newcommand{\Abstract}[2]{{\footnotesize\begin{center}ABSTRACT\end{center}
\vspace{1mm}\par#1\par
\noindent
{~}{\it #2}}}

\newcommand{\TabCap}[2]{\begin{center}\parbox[t]{#1}{\begin{center}
  \small {\spaceskip 2pt plus 1pt minus 1pt T a b l e}
  \refstepcounter{table}\thetable \\[2mm]
  \footnotesize #2 \end{center}}\end{center}}

\newcommand{\TableSep}[2]{\begin{table}[p]\vspace{#1}
\TabCap{#2}\end{table}}

\newcommand{\FigCap}[1]{\footnotesize\par\noindent Fig.\  %
  \refstepcounter{figure}\thefigure. #1\par}

\newcommand{\TableFont}{\footnotesize}
\newcommand{\TableFontIt}{\ttit}
\newcommand{\SetTableFont}[1]{\renewcommand{\TableFont}{#1}}

\newcommand{\MakeTable}[4]{\begin{table}[htb]\TabCap{#2}{#3}
  \begin{center} \TableFont \begin{tabular}{#1} #4 
  \end{tabular}\end{center}\end{table}}

\newcommand{\MakeTableSep}[4]{\begin{table}[p]\TabCap{#2}{#3}
  \begin{center} \TableFont \begin{tabular}{#1} #4 
  \end{tabular}\end{center}\end{table}}

\newenvironment{references}%
{
\footnotesize \frenchspacing
\renewcommand{\thesection}{}
\renewcommand{\in}{{\rm in }}
\renewcommand{\AA}{Astron.\ Astrophys.}
\newcommand{\AAS}{Astron.~Astrophys.~Suppl.~Ser.}
\newcommand{\ApJ}{Astrophys.\ J.}
\newcommand{\ApJS}{Astrophys.\ J.~Suppl.~Ser.}
\newcommand{\ApJL}{Astrophys.\ J.~Letters}
\newcommand{\AJ}{Astron.\ J.}
\newcommand{\IBVS}{IBVS}
\newcommand{\PASP}{P.A.S.P.}
\newcommand{\Acta}{Acta Astron.}
\newcommand{\MNRAS}{MNRAS}
\renewcommand{\and}{{\rm and }}
\section{{\rm REFERENCES}}
\sloppy \hyphenpenalty10000
\begin{list}{}{\leftmargin1cm\listparindent-1cm
\itemindent\listparindent\parsep0pt\itemsep0pt}}%
{\end{list}\vspace{2mm}}

\def\TYLDA{~}
\newlength{\DW}
\settowidth{\DW}{0}
\newcommand{\dw}{\hspace{\DW}}

\newcommand{\refitem}[5]{\item[]{#1} #2%
\def\REFARG{#3}\ifx\REFARG\TYLDA\else, {\it#3}\fi
\def\REFARG{#4}\ifx\REFARG\TYLDA\else, {\bf#4}\fi
\def\REFARG{#5}\ifx\REFARG\TYLDA\else, {#5}\fi.}

\newcommand{\Section}[1]{\section{#1}}
\newcommand{\Subsection}[1]{\subsection{#1}}
\newcommand{\Acknow}[1]{\par\vspace{5mm}{\bf Acknowledgments.} #1}
\pagestyle{myheadings}

\def\thefootnote{\fnsymbol{footnote}}

\begin{center}
{\Large\bf The Distance to the Magellanic Clouds}
\vskip1cm
{\bf
B~o~h~d~a~n~~P~a~c~z~y~{\'n}~s~k~i}
\vskip3mm
{Princeton University Observatory, Princeton, NJ 08544-1001, USA\\
e-mail: bp@astro.princeton.edu}
\end{center}

\Abstract{
The distance to LMC and SMC is a subject of controversy, with the difference
between the extreme values in distance moduli exceeding 0.5 mag.  While
currently the best calibrated method is based on red clump giants, and 
the near future improvement is most likely to come from detached eclipsing
binaries, the ultimate goal is to have a purely geometrical determination.
The best prospect will be to use relatively wide binary stars, for which 
spectroscopic orbits will be obtained with large ground based telescopes, and
astrometric orbits will be obtained either with SIM, or with future ground 
based interferometers.  A preliminary list of 25 candidate systems is 
presented.  It is based on OGLE catalogs of BVI photometry.
}{~}

\noindent
{\bf Key words:}{\it Magellanic Clouds - binaries: 
spectroscopic - binaries: visual - Stars: distances}

\Section{Introduction}

The distance to the Large Magellanic Cloud is not agreed upon.  A recent
compilation of the values of distance modulus published in 1998 and 1999
(Gibson 1999) cover a range from 18.07 to 18.74, which corresponds to
a factor 1.36 in the distance.  An attempt to harmonize the distance
scale based on RR Lyrae variables and red clump giants, the two best
calibrated indicators, gives a range 18.24 - 18.44 (Popowski 2001).
I think the current empirical calibration of the red clump giants is the
most reliable, and this gives $ 18.24 \pm 0.08 $ (Udalski 2000), but
the issue is far from being settled.

In the near future the most reliable results will be obtained with analysis
of detached eclipsing binaries.  A very good description of method, and many
historical references dating from 1910, can be found in Kruszewski and 
Semeniuk (1999).  The method was applied by Lacy (1979) to several dozens 
nearby systems, and the distances he obtained have been verified with the 
Hipparcos parallaxes (Ribas et al. 1998, Semeniuk 2000).  A much better 
calibration will be provided by the future
Full-sky Astrometric Mapping Explorer (FAME, Horner et al. 1999, Semeniuk
2001).  However, even today this method provides the LMC distance which is
competitive with the best alternatives.  The first modern determination
of the LMC distance modulus based on eclipsing binaries were by Bell et al.
(1991) and Bell et al. (1993), who used HV2226 and HV5936 to obtain
$ 18.6 \pm 0.3 $ and $ 18.1 \pm 0.3 $, respectively.  More recently
HV 2274 was observed by Guinan and his collaborators, and they obtained
for the distance modulus values $ 18.54 \pm 0.08 $, $ 18.42 \pm 0.07 $,
and $ 18.30 \pm 0.07 $ (Guinan et al. 1997, 1998a, 1998b, respectively).
The same data but different analysis of the interstellar reddening provided
somewhat different values: $ 18.22 \pm 0.13 $ (Udalski et al. 1998b) and
$ 18.40 \pm 0.07 $ (Nelson et al. 2000).  Finally, Fitzpatrick et al. (2000)
obtained $ 18.31 \pm 0.09 $ using HV982.  While the agreement between various
determinations is not perfect, they all are in a reasonable agreement 
with the `short' distance scale to the LMC.

While I expect that there will be a major improvement in the distance
determinations using detached eclipsing binaries, and the errors in the 
distance modulus will be reduced to 0.05 mag, or even less, it is far from
clear that the proponents of a different distance value will be convinced.
Ultimately, it would be important to obtain a 1\% distance using some purely
geometric method.  This will not be easy, as even if the Space Interferometry
Mission (SIM, Allen et al. 1997) achieves astrometric accuracy of $ 1 ~ \mu s $,
this will translate into 5\% accuracy for the distance determination.
While in principle the accuracy could be improved by measuring parallaxes
to many stars in the LMC, it is not likely that systematic errors will
be reduced so much as to make it practical.

In the following section I present an outline of a very traditional
and purely geometrical method for distance determination based on
a combination of astrometry and spectroscopy of visual binaries.
In the subsequent section I present a list of candidate objects,
selected from the published OGLE catalogs of BVI photometry for over 7 million
stars in the LMC (Udalski et al. 2000), and over 2 million stars in the SMC
(Udalski et al. 1998a).  Finally, I outline an observational approach to
the selection of the best systems.

A similar approach to the distance determination was proposed by Massa and
Endal (1987ab).

\Section{Geometric Distance to the Magellanic Clouds}

The simplest and best known geometrical method for distance determination
is the traditional parallax.  Unfortunately, at the LMC distance even SIM
will be able to perform 5\% measurement, at best.  Fortunately, there is
another, almost equally traditional method, based on visual binaries.
Combining astrometric and spectroscopic orbits provides 1\% distances
and masses.  Two good examples are $ \iota $ Pegasi (Boden et al. 1999a)
and 64 Piscium (Boden et al. 1999b).  This method is likely to settle
the controversy over Pleiades distance by determining the orbit of Atlas
(Pan et al. 1999).  

A great advantage of visual binary method over the
parallax is that only small angle relative astrometry is needed.  Also,
binary orbits may be much larger than 1 AU, making it possible to reach
farther with excellent accuracy.  One problem is the need to resolve the
binary, at least in the conventional use of the binary method.  This is
a serious problem at the LMC distance of $ \sim 50 $ kpc.  For a binary
in a circular orbit the angular separation is given as
$$
\theta = 0.25 ~ mas ~ \left( { P_{orb} \over 10 ~ yr } \right) ^{2/3}
\left( { M_1 + M_2 \over 20 ~ M_{\odot} } \right) ^{1/3} ,
\hskip 1.0cm {\rm for} \hskip 0.5cm {\rm d = 50 ~ kpc}.
\eqno(1)
$$
An orbital period of $ \sim 10 $ years is about as long as acceptable, 
considering rapid progress of technology.  The binary HV2274 has the
total mass of $ 23.5 ~ M_{\odot} $ Guinan et al. 1998b), and the apparent
magnitude $ V = 14.16 $ (Udalski et al. 1998b), hence the mass scaling 
adopted in the eq. (1) is reasonable.
There are two problems with an application of visual binary
method to LMC distance determination: 0.25 mas is too small an angle,
and at $ V \approx 14 $ mag a binary is too faint for current optical and
infrared interferometers.  However, it is not unreasonable to expect that
within the next 5 or 10 years such a binary will be within reach of future
ground based instruments.

There is another possibility: to use SIM do perform 1 micro-arc-second 
astrometry on the light centroid of a visual binary.  With the expected 
$ \sim 10 $ mas angular resolution SIM could not resolve the binary but 
it could determine the motion of the light centroid with an accuracy 
better than 1\%, if the amplitude of centroid motion is 0.25 mas.  To make 
this useful we have to select visual binaries made of two diverse components:
one very hot and blue, the other very cool and red, and to take advantage
of the broad spectral response of the SIM, with many filters covering 
wavelengths in the range $ 0.4 - 0.8 $ microns, approximately.  The location
of the light centroid would be close to the hot star in the blue, and close
to the cool star in the near infrared.  Of course, to be feasible, this
approach requires us to decompose the binary spectrum into two components,
and to determine the luminosity ratio of the two stars as a function of
wavelength.  While this is not an easy task, there is no obvious reason for
it to be impossible to accomplish.  In fact it should be done from the ground
prior to any attempt to put a candidate binary on the SIM target list.

Note, that Hipparcos discovered many new visual unresolved binaries 
by measuring periodic variations in the position of their light centroid
(Mignard 1998, and references therein).

For any of the two proposed approaches: either the resolution of an LMC binary
with a future ground based interferometer, or the astrometry of the light
centroid variations with the future SIM, it is necessary to identify suitable
binaries and to determine accurate spectroscopic orbits.  With the target 
stars brighter than 14 mag the determination of accurate radial velocity
curve should not be a problem.  The candidate stars are selected from the
OGLE catalogs of millions of stars in the next section.

\Section{Candidate Binaries}

Udalski et al. (1998a, 2000) published BVI photometry for over 2 million
SMC stars, and over 7 million LMC stars, respectively.  The results for the
SMC and LMC are available over the Internet at:\\
\centerline{http://bulge.princeton.edu/\~~ogle/ogle2/bvi\_maps.html}
\centerline{http://bulge.princeton.edu/\~~ogle/ogle2/lmc\_maps.html}
and at:\\
\centerline{http://www.astrouw.edu.pl/\~~ogle/ogle2/bvi\_maps.html}
\centerline{http://www.astrouw.edu.pl/\~~ogle/ogle2/lmc\_maps.html}

The stars selected for analysis were brighter than $ V = 14 $ mag, and had 
good photometry in all three bands: B, V, and I.  There were 810 such stars
in the SMC, and 1127 in the LMC.  They are shown in Fig. 1 and Fig. 2
in $ (V-I) - (B-V) $ color-color diagrams, and in Fig. 3 and Fig. 4 in
color-magnitude diagrams.  The diagonal lines in Fig. 1 and Fig. 2 correspond
to the relation:
$$
(B-V) = { 7 \over 6 } ~ (V-I) - 0.5 .
\eqno(2)
$$
The open circles located below this line are the stars which have an excess of
light in the B and I bands, i.e. these are the candidates for unresolved 
blue-red binaries.  They are all listed in Table 1, in which the second
columns give the OGLE field number, and the third columns the star number
within the corresponding fields.  All other columns are self-explanatory.
Finding charts for all the candidate binaries are shown in Fig. 5 and Fig. 6.
The candidate is located at the center of each square panel, which is 1,000
pixels on its side, corresponding to 7' in the sky.

\Section{Discussion}

The candidates listed in Table 1 have to be verified spectroscopically for 
the presence of composite spectra.  The next step is more time consuming:
for every star with a composite spectrum the two radial velocity curves 
have to be determined.  Some genuine binaries will have periods too short, 
and hence angular separations too small to be of interest.  Others
will have periods too long to be of interest.  Hopefully, there will be
several objects in the period range of 5 -- 15 years, which might be 
considered to be optimal for distance determination.

The list of candidate binaries will increase as OGLE upgrades to a large
mosaic CCD camera, and multi-band photometry will become available for
full SMC and LMC, not just for their central regions.  Also, photometry for
the stars which are brighter than the saturation limit apparent in Fig. 3 
and Fig. 4 should be searched for possible candidates.

\Acknow{This work was supported with the NSF grants AST-9819787
and AST-9820314.  It is a great pleasure to acknowledge that all figures
were made using the SM plotting package provided by R. H. Lupton.}


\newpage

\begin{figure}[htb]
\psfig{figure=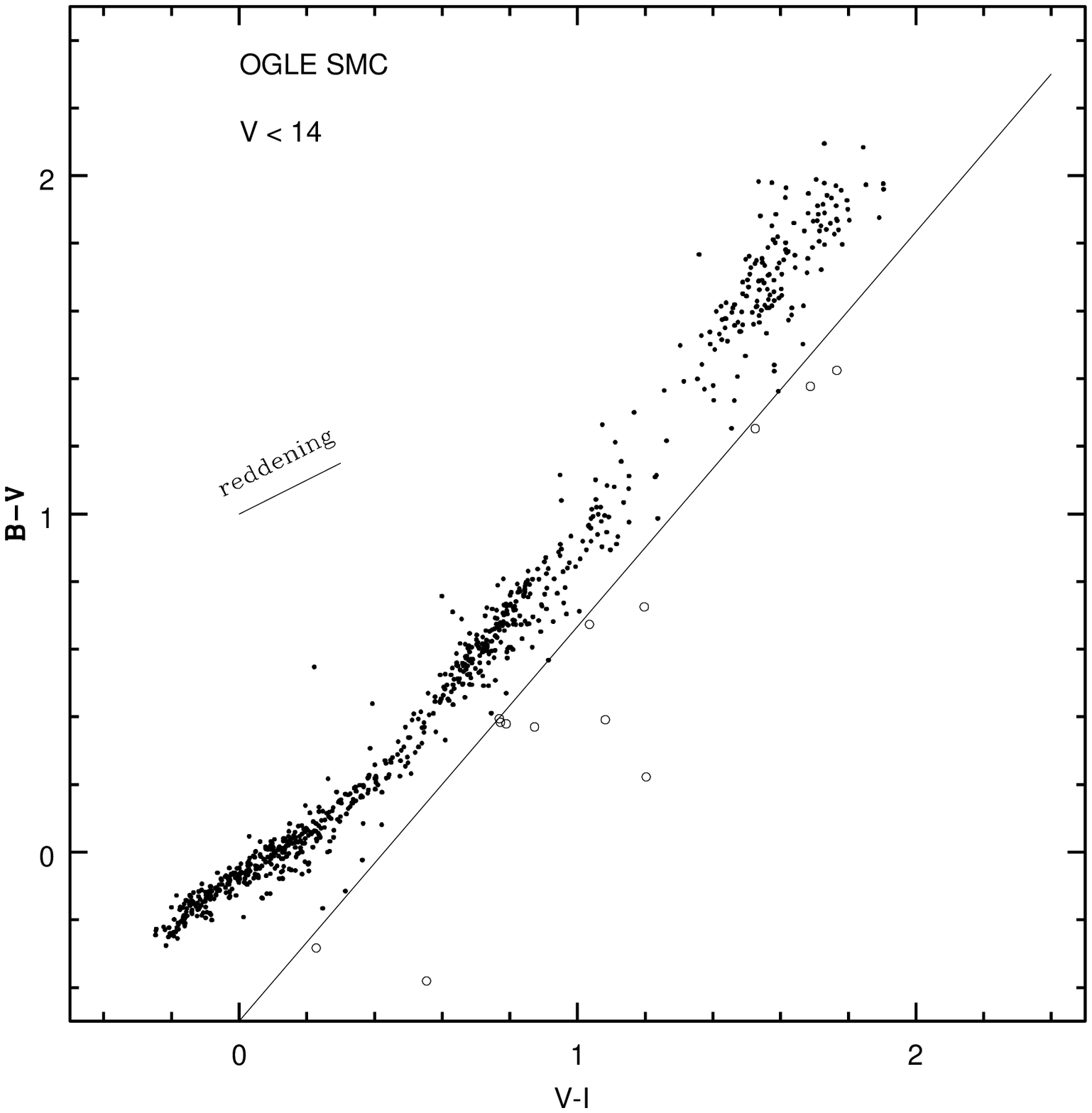,bbllx=20pt,bblly=150pt,bburx=550pt,bbury=680pt,width=12cm,
clip=}
\vspace*{3pt}
\FigCap{Color-color diagram for 810 OGLE stars in the SMC that are 
brighter than $ V = 14 $ mag.  The diagonal line is given by the eq. (2).
The stars indicated with open circles have an excess of B and I light, and are
candidates for blue-red unresolved binaries.  A reddening vector is shown.
}
\end{figure}

\newpage

\begin{figure}[htb]
\psfig{figure=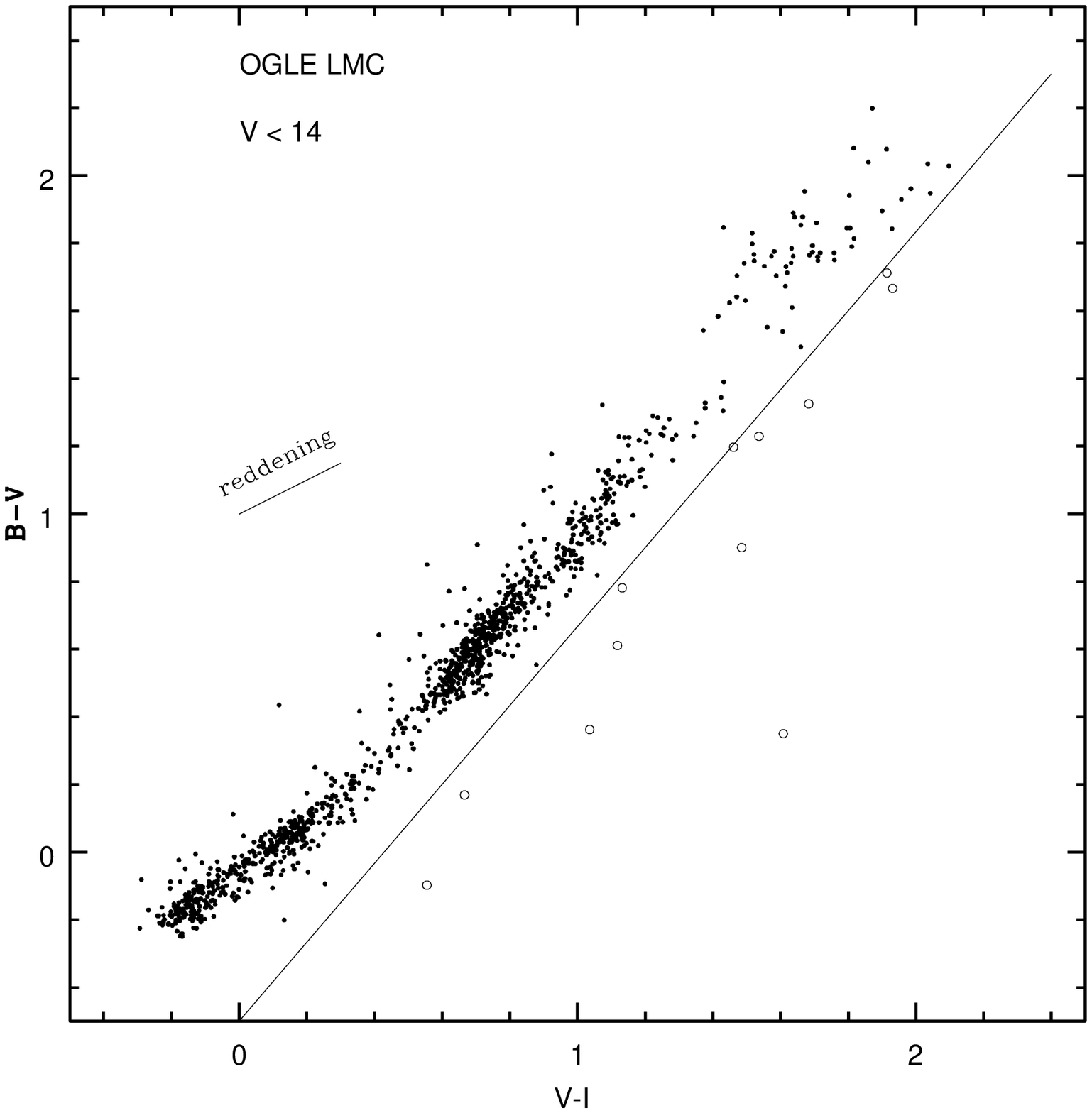,bbllx=20pt,bblly=150pt,bburx=550pt,bbury=680pt,width=12cm,
clip=}
\vspace*{3pt}
\FigCap{Color-color diagram for 1127 OGLE stars in the LMC that are 
brighter than $ V = 14 $ mag.  The diagonal line is given by the eq. (2).
The stars indicated with open circles have an excess of B and I light, and are
candidates for blue-red unresolved binaries.  A reddening vector is shown.
}
\end{figure}

\newpage

\begin{figure}[htb]
\psfig{figure=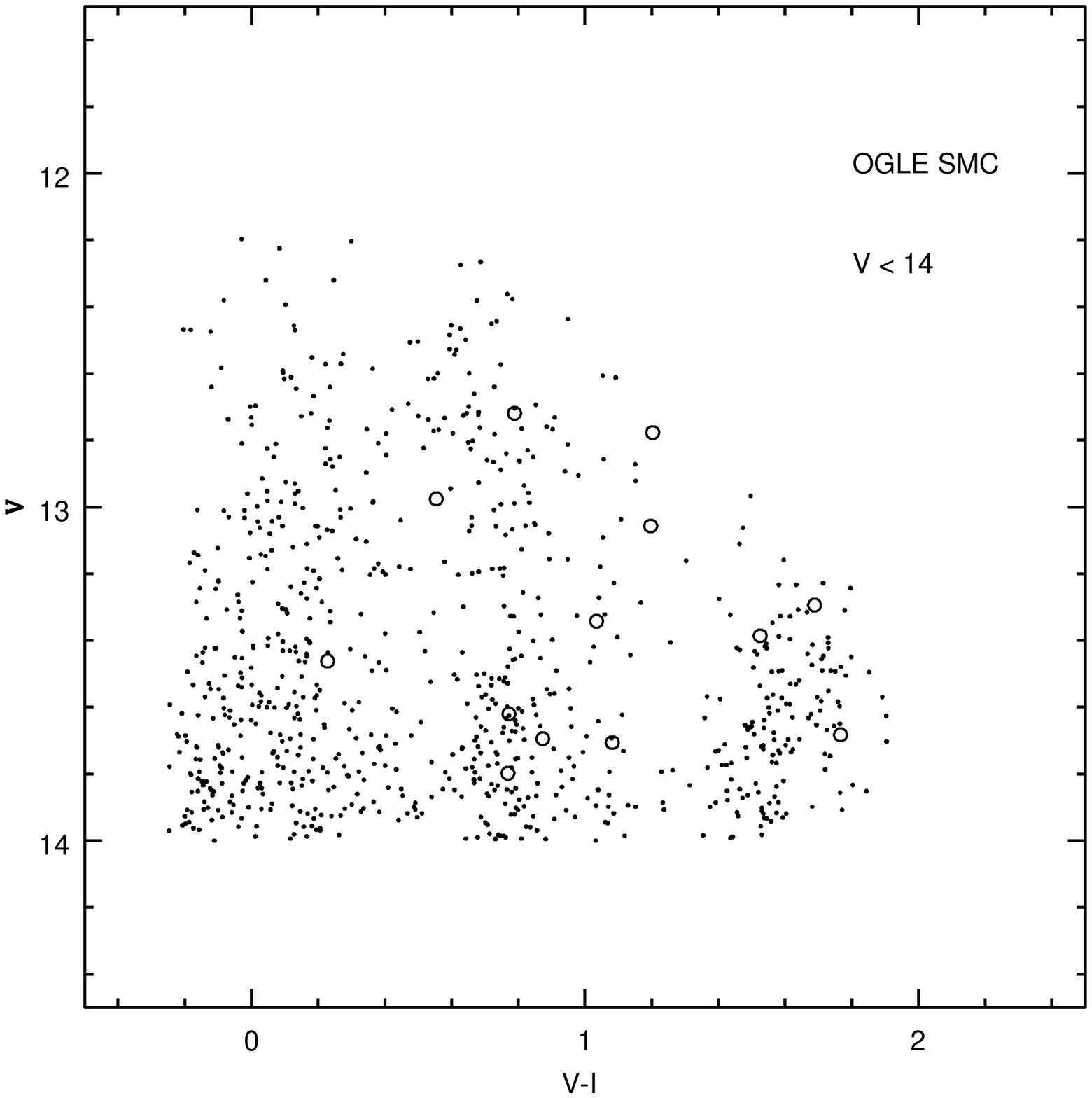,bbllx=20pt,bblly=150pt,bburx=550pt,bbury=680pt,width=12cm,
clip=}
\vspace*{3pt}
\FigCap{Color-magnitude diagram for 810 OGLE stars in the SMC that are 
brighter than $ V = 14 $ mag.  The stars
indicated with open circles have an excess of B and I light, and are
candidates for blue-red unresolved binaries.
}
\end{figure}

\newpage

\begin{figure}[htb]
\psfig{figure=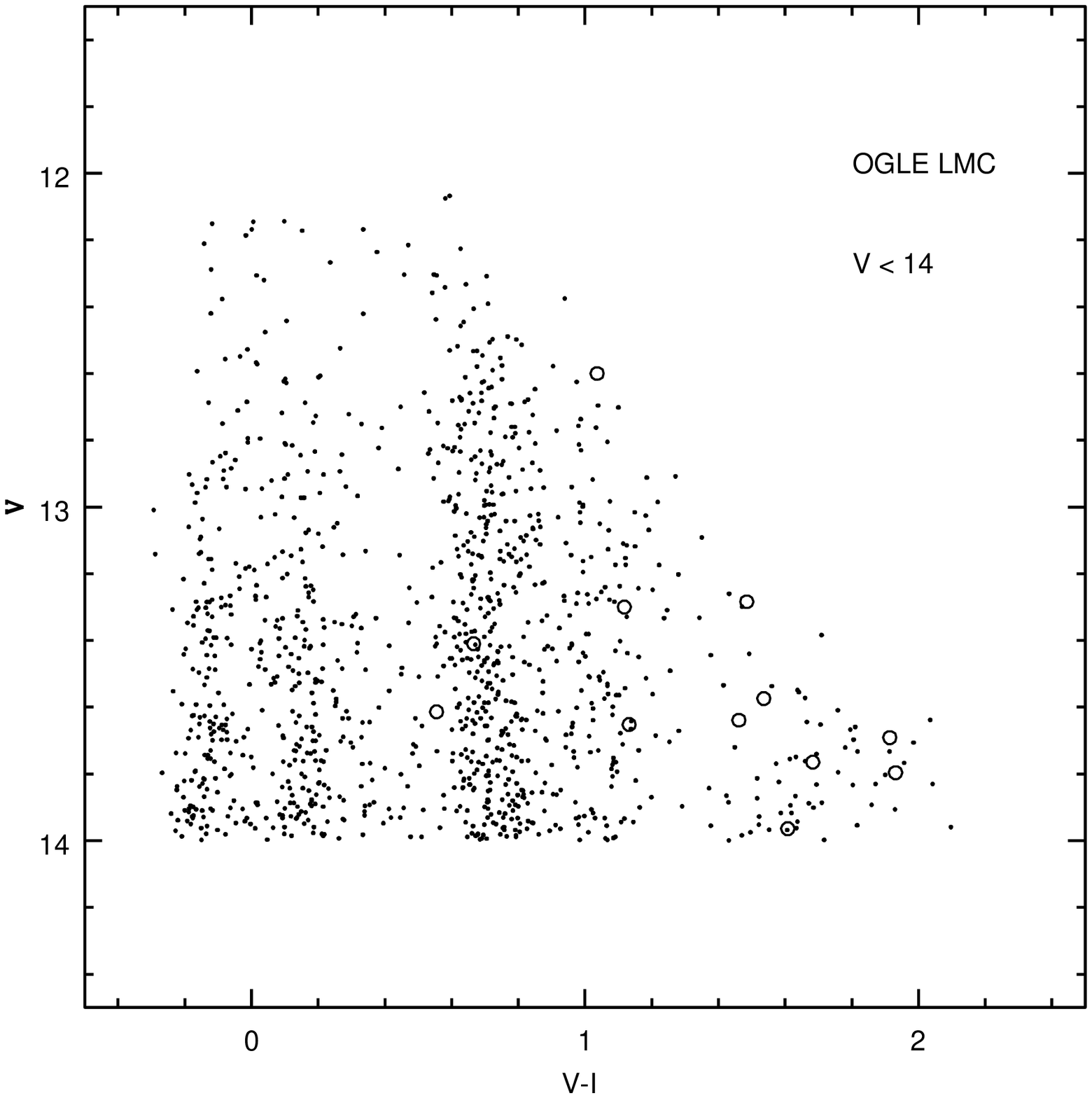,bbllx=20pt,bblly=150pt,bburx=550pt,bbury=680pt,width=12cm,
clip=}
\vspace*{3pt}
\FigCap{Color-magnitude diagram for 1127 OGLE stars in the LMC that are 
brighter than $ V = 14 $ mag.  The stars
indicated with open circles have an excess of B and I light, and are
candidates for blue-red unresolved binaries.
}
\end{figure}

\newpage

\begin{figure}[htb]
\psfig{figure=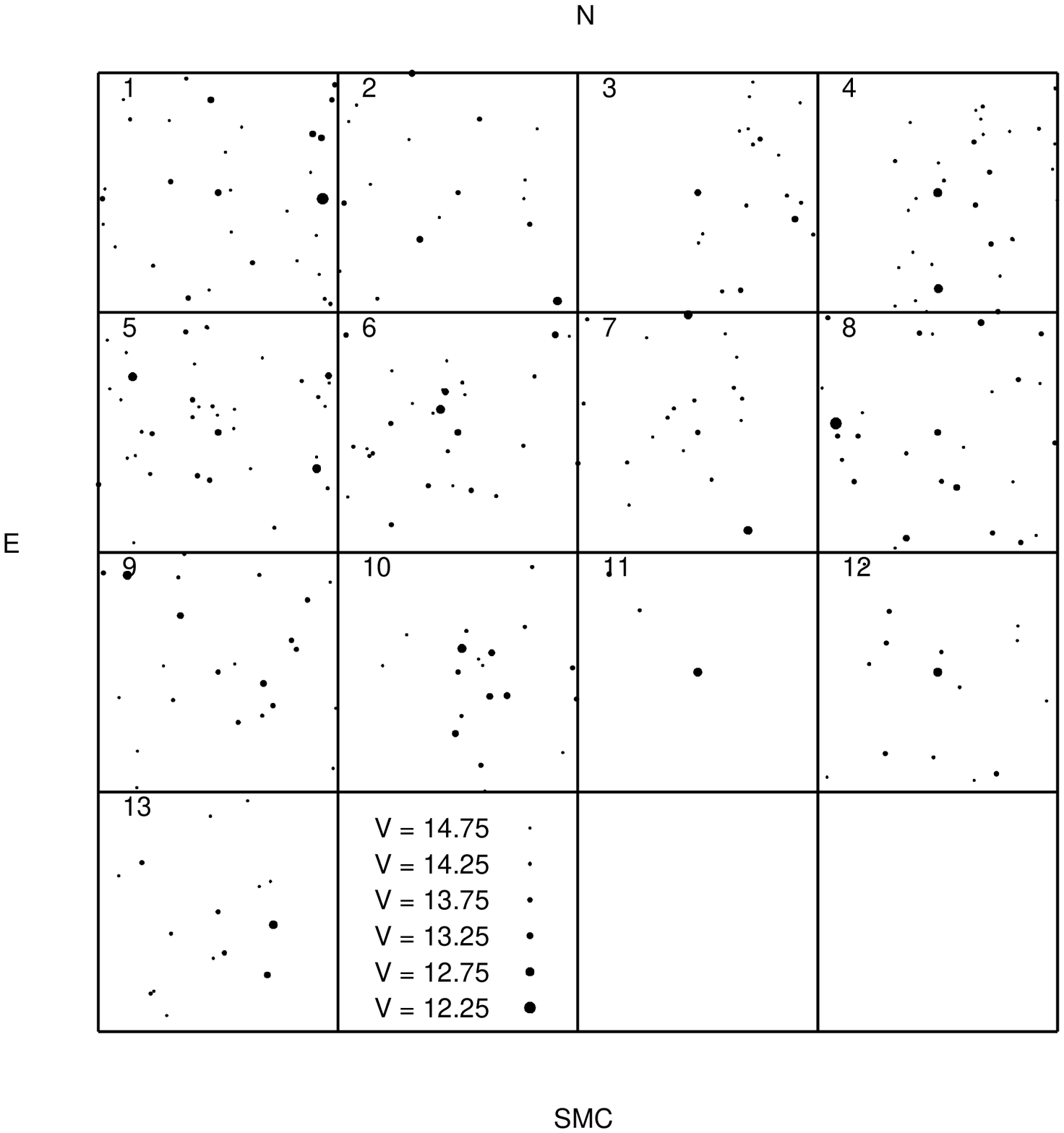,bbllx=20pt,bblly=150pt,bburx=550pt,bbury=680pt,width=12cm,
clip=}
\vspace*{3pt}
\FigCap{Finding charts for 13 candidate blue-red unresolved binaries in the 
SMC.  Each square is 7' on a side.
}
\end{figure}

\newpage

\begin{figure}[htb]
\psfig{figure=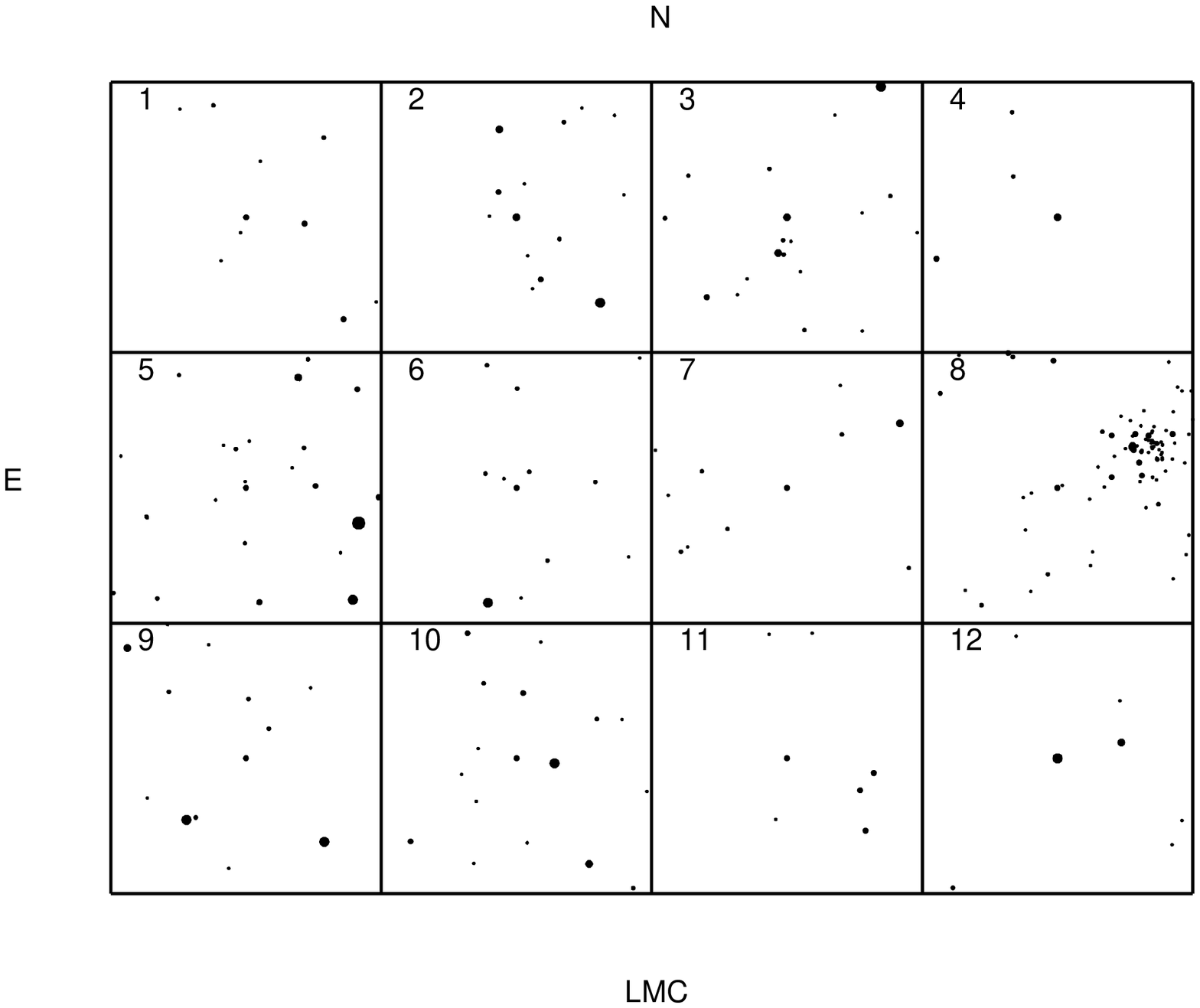,bbllx=20pt,bblly=150pt,bburx=550pt,bbury=680pt,width=12cm,
clip=}
\vspace*{3pt}
\FigCap{Finding charts for 12 candidate blue-red unresolved binaries in the 
LMC.  Each square is 7' on a side.
}
\end{figure}

\MakeTableSep{lrclllrrc@{\hspace{3pt}}cccccc}{12.5cm}{Candidate binaries in the SMC}
{\hline
\noalign{\vskip4pt}
\multicolumn{1}{c}{} & \multicolumn{1}{c}{field}   & \multicolumn{1}{c}{number} & 
\multicolumn{1}{c}{RA} & \multicolumn{1}{c}{DEC} & \multicolumn{1}{c}{V} &
\multicolumn{1}{c}{B-V} & \multicolumn{1}{c}{V-I} \\
\noalign{\vskip4pt}
\hline
\noalign{\vskip4pt}
 1  &  4 & 16915 & 0:47:08.72 & -73:14:11.5 & 13.057 &  0.725 & 1.197 \\
 2  &  5 & 19933 & 0:50:32.09 & -72:52:09.3 & 13.695 &  0.370 & 0.873 \\
 3  &  6 & 17257 & 0:51:22.85 & -73:15:41.8 & 13.462 & -0.283 & 0.228 \\
 4  &  6 & 35446 & 0:51:45.43 & -73:04:59.7 & 12.719 &  0.379 & 0.789 \\
 5  &  6 & 10522 & 0:52:19.06 & -73:09:22.7 & 13.342 &  0.673 & 1.035 \\
 6  &  6 & 50956 & 0:53:44.49 & -72:33:19.0 & 13.294 &  1.377 & 1.688 \\
 7  &  6 & 11169 & 0:54:09.53 & -72:41:42.9 & 13.706 &  0.392 & 1.082 \\
 8  &  9 & 14255 & 1:02:37.90 & -72:35:54.4 & 13.386 &  1.252 & 1.525 \\
 9  & 10 & ~~545 & 1:05:10.31 & -72:16:44.4 & 13.683 &  1.425 & 1.766 \\
10  & 11 & 53227 & 1:07:04.55 & -72:25:51.8 & 13.798 &  0.395 & 0.769 \\
11  & 11 & 63522 & 1:08:30.22 & -73:05:13.3 & 12.777 &  0.222 & 1.203 \\
12  & 11 & 77218 & 1:07:54.31 & -72:40:58.7 & 12.976 & -0.381 & 0.554 \\
13  & 11 & 15501 & 1:08:54.79 & -72:24:52.0 & 13.620 &  0.384 & 0.772 \\
\hline}
\MakeTableSep{lrclllrrc@{\hspace{3pt}}cccccc}{12.5cm}{Candidate binaries in the LMC}
{\hline
\noalign{\vskip4pt}
\multicolumn{1}{c}{} & \multicolumn{1}{c}{field}   & \multicolumn{1}{c}{number} & 
\multicolumn{1}{c}{RA} & \multicolumn{1}{c}{DEC} & \multicolumn{1}{c}{V} &
\multicolumn{1}{c}{B-V} & \multicolumn{1}{c}{V-I} \\
\noalign{\vskip4pt}
\hline
\noalign{\vskip4pt}
 1  &  3 & 79892 & 5:27:54.83 & -69:39:32.6 & 13.614 & -0.098 & 0.555 \\
 2  &  4 & 52814 & 5:26:39.03 & -69:25:25.4 & 13.300 &  0.611 & 1.118 \\
 3  &  6 & 14228 & 5:20:39.59 & -69:19:31.2 & 13.284 &  0.900 & 1.485 \\
 4  &  6 & 69953 & 5:22:39.40 & -69:57:12.3 & 13.410 &  0.169 & 0.666 \\
 5  &  7 & 95146 & 5:19:24.33 & -69:25:45.9 & 13.765 &  1.326 & 1.683 \\
 6  &  8 & 15413 & 5:15:06.94 & -69:35:54.2 & 13.964 &  0.350 & 1.608 \\
 7  &  9 & 88658 & 5:14:02.12 & -69:39:56.1 & 13.691 &  1.712 & 1.914 \\
 8  & 11 & 62232 & 5:08:18.31 & -68:46:47.2 & 13.652 &  0.782 & 1.132 \\
 9  & 11 & 13943 & 5:09:36.47 & -69:04:54.7 & 13.796 &  1.666 & 1.931 \\
10  & 14 & 50495 & 5:03:02.40 & -68:47:20.5 & 13.639 &  1.197 & 1.461 \\
11  & 15 & ~~~~2 & 5:00:03.03 & -69:30:08.6 & 13.574 &  1.229 & 1.536 \\
12  & 19 & 89862 & 5:43:12.33 & -70:17:32.3 & 12.600 &  0.362 & 1.036 \\
\hline}

\end{document}